\documentclass[runningheads,a4paper]{llncs}

\usepackage{amssymb}
\usepackage{graphicx}
\usepackage{epsfig}
\usepackage{amsmath}
\usepackage{amsfonts}
\usepackage{arydshln}
\usepackage{url}
\usepackage{graphicx}
\usepackage{lipsum}
\usepackage{epsfig}
\usepackage{graphicx,epstopdf}
\DeclareGraphicsExtensions{.ps}
\usepackage{verbatim}
\usepackage{multicol}
\usepackage{subfigure}

\urldef{\mailsa}\path|{alfred.hofmann, ursula.barth, ingrid.haas, frank.holzwarth,|
\urldef{\mailsb}\path|anna.kramer, leonie.kunz, christine.reiss, nicole.sator,|
\urldef{\mailsc}\path|erika.siebert-cole, peter.strasser, lncs}@springer.com|    
\newcommand{\keywords}[1]{\par\addvspace\baselineskip
\noindent\keywordname\enspace\ignorespaces#1}
\begin{document}

\mainmatter  

\title{Batch Incremental Shared Nearest Neighbor Density Based Clustering Algorithm for Dynamic datasets}

\titlerunning{BISD Clustering Algorithm for Dynamic datasets}

%
%
\author{Panthadeep Bhattacharjee, Amit Awekar}
\authorrunning{Panthadeep Bhattacharjee, Amit Awekar}

\institute{Department of Computer Science \& Engineering\\Indian Institute of Technology, Guwahati, India\\
Email: \{panthadeep,awekar\}@iitg.ernet.in}

%
%

\toctitle{Batch Incremental Shared Nearest Neighbor Density Based Clustering Algorithm for Dynamic datasets}
\tocauthor{Panthadeep Bhattacharjee}
\maketitle

\begin{abstract}
Incremental data mining algorithms process frequent updates to dynamic datasets efficiently by avoiding redundant computation.  Existing incremental extension to shared nearest neighbor density based clustering (SNND)  algorithm cannot handle deletions to dataset and handles insertions only one point at a time. We present an incremental algorithm to overcome both these bottlenecks by efficiently identifying affected parts of clusters while processing updates to dataset in batch mode. We show effectiveness of our algorithm by performing experiments on large synthetic as well as real world datasets. Our algorithm is up to four orders of magnitude faster than SNND and requires up to 60\% extra memory than SNND while providing output identical to SNND.
\keywords{Clustering, Dynamic datasets}
\end{abstract}

\section{Introduction}
Many popular clustering algorithms work on a static snapshot of dataset. However, many real-world applications such as search engines and recommender systems are expected to work over dynamic datasets. Such datasets undergo frequent changes where some points are added and some points are deleted. A naive method to get exact clustering over the changed dataset is to run the clustering algorithm again. Most of the computation in reclustering is going to be redudant. This problem becomes more severe with increase in frequency of updates to dynamic datasets. For large datasets, rerun of the algorithm might not finish before next batch of updates arrive. Incremental algorithms target this fundamental issue of redundant computation yet obtain output identical to their non-incremental counterpart. Incremental clustering problem can be defined more formally as follows. Consider a data set $D$ alongwith its initial clustering $f : D \rightarrow \mathbb{N} $ and a sequence of $n$ updates that convert $D$ to $D'$. An incremental clustering is defined as a mapping $g : f,D' \rightarrow \mathbb{N} $ isomorphic to the one-time clustering $f(D')$ by the non-incremental algorithm.

SNND\cite{ertoz2003finding} is a widely used clustering algorithm for its robustness while finding clusters of varying densities and shapes even among high dimensional data. SNND converts a given data set to the SNN graph and clusters this graph. Efficiently computing changes to the SNN graph is the key challenge in designing an incremental extension to SNND. Existing incremental extension for SNND known as Incremental Shared nearest neighbor Density-based clustering (InSD) \cite{singh2013incremental} has two main bottlenecks: First, it does not support deletion of points. Second, it supports addition of only one point at a time. If multiple updates are made to the dataset then, InSD treats each update independently and reclusters the dataset in response to each individual update. With increasing size of updates, InSD is only marginally faster than SNND and in some cases even slower.

Our goal is to design a Batch Incremental Shared Nearest Neighbor Density based Clustering (BISD) algorithm that overcomes both the bottlenecks of InSD and satisfies all three following conditions: 1. $T_{BISD} \leq T_{InSD}$,  2. $T_{BISD} \leq T_{SNND}$, and 3. $C_{BISD} = C_{SNND}$. Here $T$ indicates running time of the algorithm and $C$ indicates clustering provided by the algorithm. Unlike InSD, our algorithm integrates effect of multiple changes to the dataset by computing updates to the SNN graph in single execution. BISD achieves significant speedup at the cost of additional memory footprint while guaranteeing the same output. We show effectiveness of BISD by performing experiments over synthetic and real world datasets. Our datasets range in dimensionality from 2 to 70 and the number of points vary from 13000 to 100000. We observe that BISD is consistently multiple orders of magnitude faster than SNND while producing exact same output. BISD always maintains speed up over InSD.


\section{Related work}\label{RelWork}
Density based clustering paradigm was introduced by Sanders et. al. in their DBSCAN algorithm\cite{sander1998density}. It computes density for each point in terms of number of points that are within certain distance. However, computing density using distance does not work well for high dimensional data and clusters with varying density. Shared nearest neighbor (SNN) based clustering paradigm was introduced by Jarvis \& Patrick\cite{Jarvis_Patrick}. Their algorithm computes top $k$ nearest neighbors (KNN) list for each point. Similarity between points is the extent of overlap between their corresponding KNN lists. Dataset is represented as SNN graph, which is a weighted, undirected, simple graph. Each vertex of the graph is a point. Edge exists between two vertices if they are in each other's KNN list. Weight of the edge is the similarity score computed as mentioned before. Edges with weight below the similarity threshold ($\delta_{sim}$) are deleted from the SNN graph. Each component of the remaining graph is treated as a cluster.

SNND\cite{ertoz2003finding} algorithm combines two paradigms: density and SNN. It creates SNN graph similar to Jarvis \& Patrick algorithm. However, it does not settle with simple idea of components as clusters. A vertex is labeled as core if its degree in SNN graph is above the strong link formation threshold ($\delta_{core}$). A vertex is an outlier if it is neither core nor connected to any core vertex. Such vertices are not considered as part of any cluster. Rest of the vertices are labeled as non-core. Initially each core vertex is considered as a separate cluster. Two clusters are merged if at least one core vertex from one cluster is connected with a core vertex in other cluster. This process is repeated till no more clusters can be merged. A non-core vertex is assigned to a cluster having strongest link with it.


InSD\cite{singh2013incremental} is an incremental extension of SNND. It does not handle deletion of points. InSD handles insertion of points only one point at a time. When a new point is added to the dataset, its KNN list is computed. This new point might occupy place in KNN lists of existing points, resulting in addition and deletion of edges from the SNN graph.  InSD first goes through merge phase to identify clusters that will merge based on newly added edges to the SNN graph. Then InSD goes through split phase to identify clusters that will split into multiple fragments based on edges removed from the SNN graph. Output of InSD is identical to SNND. If multiple points are inserted in the dataset then, InSD processes each insertion independently and reclusters the whole dataset before handling next insertion.


\section{Proposed Algorithm} \label{OurWork}

Given a dataset $D$, initial clustering $f$, a sequence of $n$ changes that converts $D$ to $D'$, extended nearest neighbors list for each point in D and SNN graph for D, our BISD algorithm provides clustering of $D'$ that is identical to SNND. BISD partitions the $n$ changes into set of points to be inserted ($n_{add}$) and deleted ($n_{del}$). First it handles all insertions, followed by all deletions to compute changes to the SNN graph and reclusters the data in single execution. BISD algorithm needs four parameters: $\delta_{sim}$, $\delta_{core}$, $k$, and $w$. First three parameters are identical to parameters used by SNND and InSD (Please refer Section \ref{RelWork}). Fourth parameter $w$ is the size of the extended KNN list ($w\geq k$). For each point BISD computes $w$ nearest neighbors, but uses only top $k$ out of them for clustering. Top k nearest neighbor list for a point P is denoted as $k(P)$ and its size as $k$. Extended nearest neighbor list for point $P$ is denoted as $w(P)$ and its size as $w$.  $k(P)$ is not stored separately. It is generated using top k elements of $w(P)$.
   
\subsection{Insertion phase} \label{InsertionPhase}
Insertion phase of BISD algorithm converts $D$ to $D'_{add}$ by adding all points in $n_{add}$. For each point $P$ in $n_{add}$, $w(P)$ and $k(P)$ lists are created by computing distance with every point in D and every other point in $n_{add}$. Points in original dataset $D$ are partitioned into three sets. A point in $D$ is added to set $T1_{add}$, if any point in $n_{add}$ becomes member of its KNN list. A point in D is added to set $T2_{add}$, if it is not already in $T1_{add}$ and at least one member of its KNN list is in $T1_{add}$. Rest of the points in D are unaffected by addition of new points. Such points are added to set $U_{add}$.

\begin{table}[!htp]
\centering
	\caption{Example point}
	\label{ExampleTable}

	\begin{tabular}{|l|l|l|l|l|l|l|l|l|}
	\hline
	Point & $P2$ & $P3$ & $P4$ & $P5$ & $N1$ & $N2$ & $N3$ & $N4$\\ \hline
	Distance from $P1$ & 5 & 10 & 20 & 30 & 40 & 25 & 7 & 2\\ \hline

	\end{tabular}

	\end{table}
		
For example, please refer to Table \ref{ExampleTable}. $P1$ is a point in $D$. Here size of $k$ is two and size of $w$ is four. Initially, $w(P1)=\{P2, P3, P4, P5\}$ and $k(P1)=\{P2, P3\}$. Let $n_{add}$ consist of four points $N1$, $N2$, $N3$, and $N4$. After adding these four points, $w(P1)=\{N4, P2, P3, N2\}$ and $k(P1)=\{N4, P2\}$. Both $w(P1)$ and $k(P1)$ change. BISD will add $P1$ to $T1_{add}$ set. However, if $n_{add}$ consisted of only points $N1$ and $N2$, then $w(P1)=\{P2, P3, P4, N2\}$ and $k(P1)=\{P2, P3\}$. Insertion of these two points affects $w(P1)$ but not $k(P1)$. In such case, $P1$ will not be added to $T1_{add}$ set.

\subsection{Deletion Phase} \label{DeletionPhase}
During the deletion phase, points listed in $n_{del}$ are removed from $D'_{add}$ to get final dataset $D'$. For each point $P$  in $D'$, $w(P)$ and $k(P)$ are updated to remove any member that belongs to $n_{del}$. If size of $w(P)$ falls below the required size of KNN list ($k$), then $w(P)$ is recalculated for $P$. Otherwise, BIDS maintains truncated $w(P)$ as only top k entries in $w(P)$ are used to generate $k(P)$ and the SNN graph. The dataset $D'$ is partitioned into three sets. Set $T1_{del}$ contains each point $P$ in $D'$ such that $k(P)$ was changed because of deletion of points. Set $T2_{del}$ contains each point $P$ in $D'$ such that $P$ does not belong to set $T1_{del}$ and at least one member of $k(P)$ belongs to $T1_{del}$. Rest of the points in $D'$ are added to the set $U_{del}$.

Let us continue the example in Section\ref{InsertionPhase}. At the end of insertion phase (after adding four points), $k(P1)=\{N4, P2\}$ and $w(P1)=\{N4, P2, P3, N2\}$. Consider first case where $n_{del}$ consists of three points $N2$, $P2$, and $P3$. After removing these three points, size of $w(P1)$ is reduced to one. It is less than the required size of $k(P1)$. In such case, $w(P1)$ is recomputed and $P1$ is added to the set $T1_{del}$. Consider second case that $n_{del}$ consists of only two points $N2$ and $P2$. In such scenario, $P1$ will still be added to set $T1_{del}$ but $w(P1)$ will not be recalculated as truncated $w(P1)$ is enough to generate $k(P1)$. Consider third case where $n_{del}$ consists of only one point $P3$. In this case, $P1$ will not be added to set $T1_{del}$ as the deletion affects $w(P1)$ but $k(P1)$ remains unchanged.

\subsection{Clustering Phase} \label{ClusteringPhase}
Clustering phase aggregates changes from insertion and deletion phases to build updated SNN graph efficiently. First it builds set $T1=(T1_{add}\cup T1_{del})-n_{del}$. $T1$ set contains each point $P$ that belongs to $D'$ such that $k(P)$ was updated either in insertion or deletion phase. Then it builds set $T2=((T2_{add}\cup T2_{del})-T1)-n_{del}$. This set contains each point $R$ that belongs to $D'$ such that $k(R)$ remained unchanged in both insertion and deletion phases, but at least one member of $k(R)$ belongs to $T1$. Points in $D'$ that are not in $T1$ as well as $T2$, are unaffected by changes in the dataset.

We represent SNN graph using adjacency lists. For each point in $n_{add}$, new vertex is added to the SNN graph. Vertices corresponding to each point in $n_{del}$ are removed from the SNN graph. For each point $P$ in $T1$, its similarity score with every member of its $k(P)$ can potentially change. Therefore the adjacency lists of vertices in $T1$ are completely updated. However, for each point $R$ in $T2$, its similarity score with a member of $k(R)$ can change only if that member is in $T1$. Hence, we only selectively update adjacency lists of vertices in $T2$. Points that are in $T1$ or $T2$ can possibly form new edges, update weights for some of the existing edges and lose some other existing edges. We can recluster this updated SNN graph using the clustering approach of $SNND$. After clustering, BISD stores the whole SNN graph, extended nearest neighbor list and cluster membership for all points to the secondary storage. This information is used during next invocation of BISD to handle new changes.
\vspace*{-\baselineskip}
	\begin{table*}[t]
	\centering
	\caption{Datasets used}
	\label{DatasetTable}
	\begin{tabular}{|l|l|l|l|}
	\hline
	Dataset & Points & Dimensions & Description \\ \hline
	Mopsi12 & 13000 & 2 & Locations in Finland \\ \hline
	5D & 100000 & 5 &  Synthetic dataset  \\ \hline
	Birch3 & 100000 & 2 & 10 by 10 grid of gaussian clusters \\ \hline
	KDDCup'04 & 60000 & 70 & Identifying homologous proteins to native sequence \\ \hline
	KDDCup'99 & 54000 & 41 &  Network intrusion detection\\  \hline

	\end{tabular}
	\end{table*}


\begin{figure*}[b]
\centering
\label{Speedup}
\resizebox{\columnwidth}{!}{
\begin{tabular}{ll}
	\subfigure[\label{SpeedSAdd}]{
			\includegraphics[scale=.6]{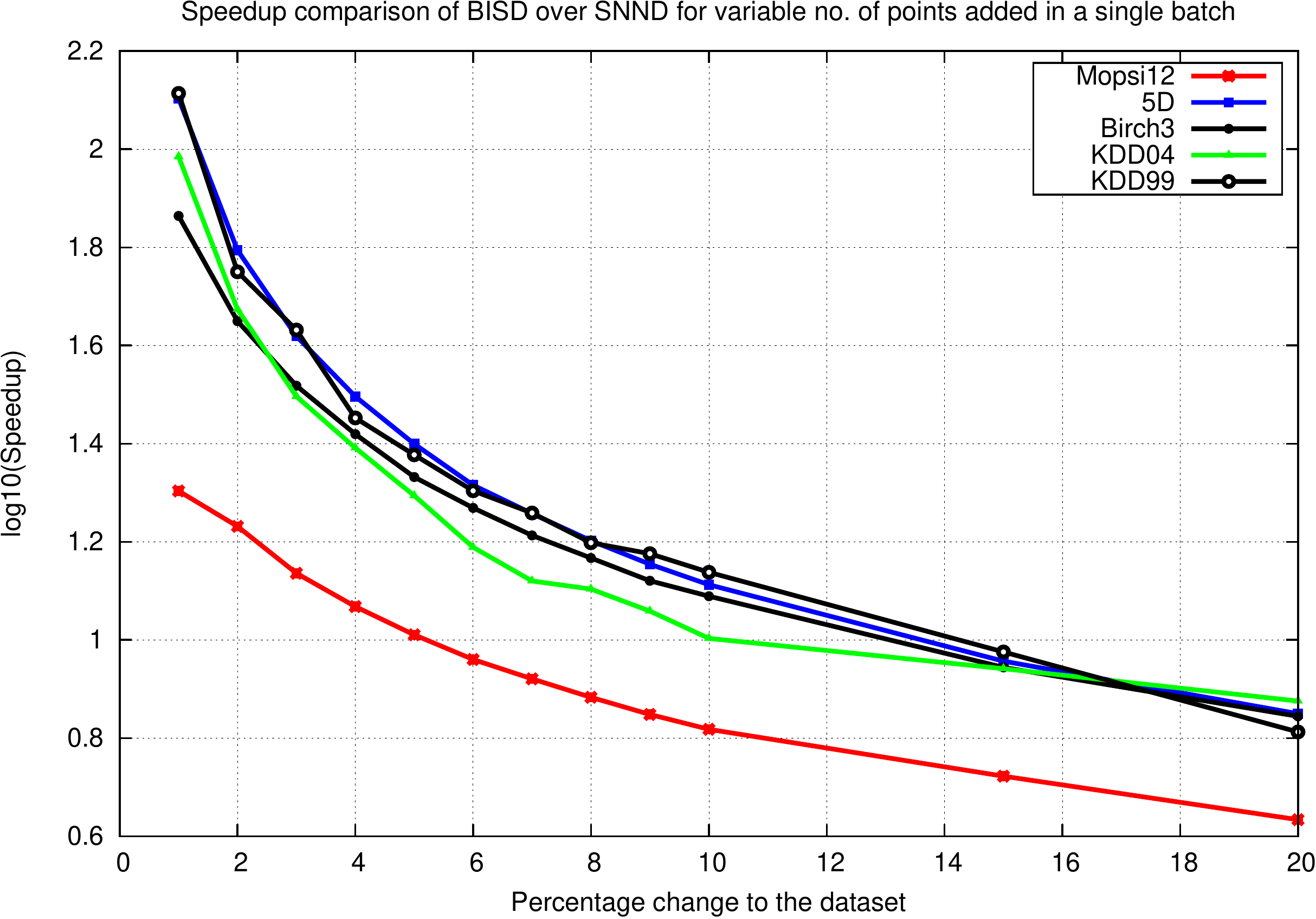}
		    }  & \subfigure[\label{SpeedInAdd}]{		
		    		\includegraphics[scale=.6]{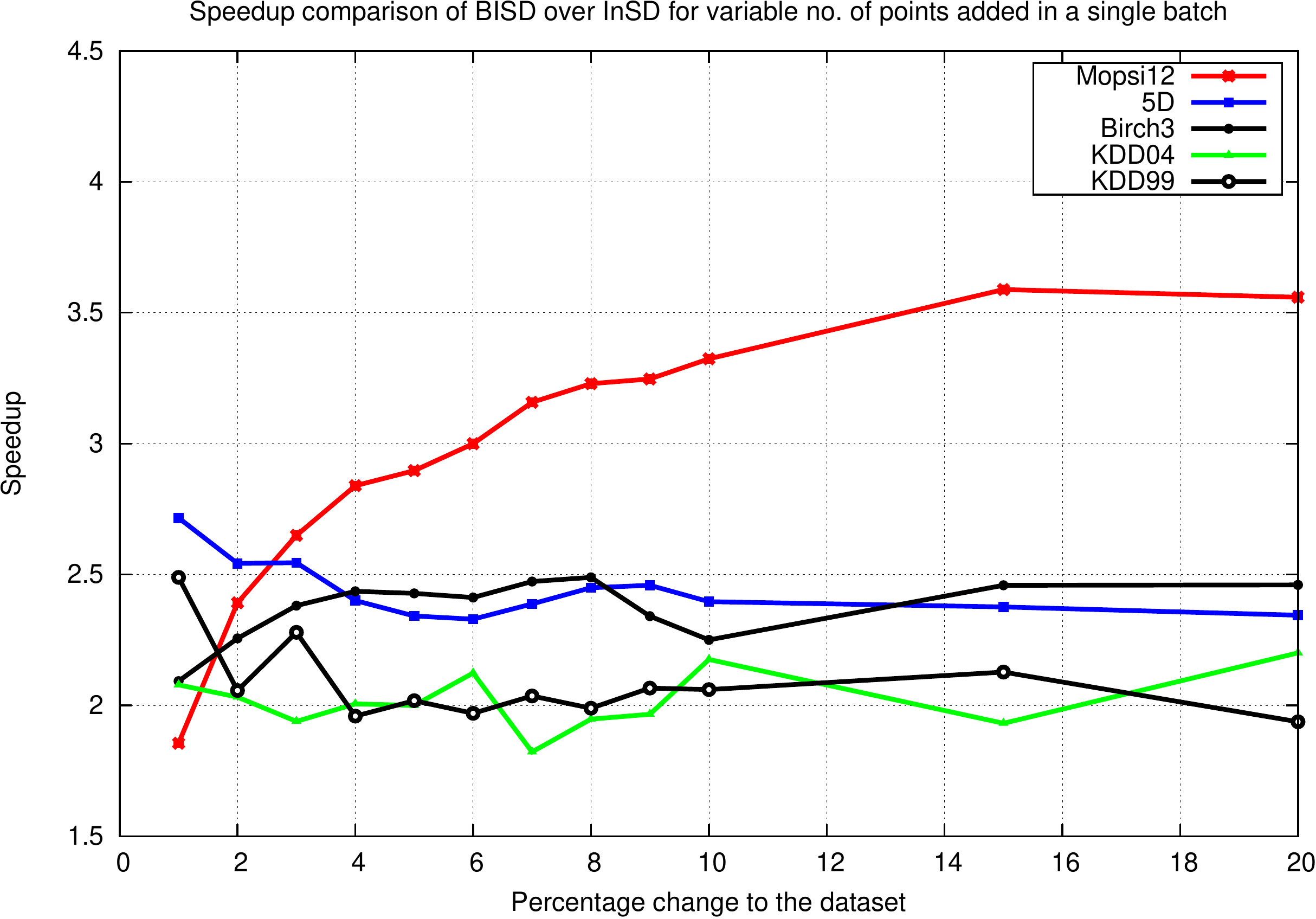}
		    	    }

\end{tabular}
}
\caption{Speedup of BISD algorithm over SNND and InSD}
\end{figure*}


\subsection{Differences from SNND and InSD}
Consider the running example mentioned in Sections \ref{InsertionPhase} and \ref{DeletionPhase}. SNND will rerun whole computation over the complete dataset. SNND will ignore the fact that only few points are affected by these changes. SNND will compute top k nearest neighbors list for every point in $D'$, build SNN graph from scratch and recluster this graph. As a result, SNND runs multiple orders of magnitude slower than BISD. However, BISD algorithm more efficiently rebuilds the SNN graph identical to SNND. Therefore clustering of BISD algorithm is identical to SNND.

	   

	   
InSD algorithm cannot handle deletion of three points. However to handle addition of four points, InSD will treat insertion of each point independently. InSD will recompute changes to SNN graph after insertion of each point and recluster the SNN graph before processing next point. InSD will not aggregate changes from multiple insertions and it will modify KNN lists more often than BISD algorithm. Therefore, InSD algorithm always runs slower than BISD.

\section{Experiments} \label{Experiments}
To demonstrate effectiveness of BISD algorithm, we conducted experiments on three real-world (Mopsi12, KDDCup'99, and KDDCup'04) and two synthetic (Birch3 and 5D) datasets. Please refer to Table \ref{DatasetTable}. All algorithms were implemented in C++ on Linux platform. Experiments were conducted on a machine with 32 GB RAM and 1.56 GHz CPU. All code and datasets are available publicly for download\footnote{https://sourceforge.net/projects/bisd/}.


Figures \ref{SpeedSAdd} and \ref{SpeedInAdd} show speed up of BISD algorithm over SNND and InSD respectively for all five datasets. X axis represents number of changes made to dataset as percentage of base dataset size. Y axis represents speed up of BISD. For Figure \ref{SpeedSAdd}, Y axis is log scaled to base ten. Speed up of BISD over SNND and InSD is calculated as $T_{SNND}/T_{BISD}$ and $T_{InSD}/T_{BISD}$ respectively. As number of changes to the dataset increase, insertion and deletion  phases of BISD take more time. This happens because number of affected points either in sets $T1$ or $T2$ also increase with more number of changes. As a result, speed up of BISD over SNND reduces with increase in number of changes. BISD maintains extended KNN list in memory for all points. This memory overhead is up to 60\% extra for BISD as compared to memory footprint of SNND. BISD achieves speed up of up to 4 over InSD as it aggregates changes over all insertions.





 \section{Conclusion and Future Work}\label{Conclusion}
We presented BISD algorithm that handles both addition and deletion to the dataset. It aggregates changes to the SNN graph across multiple changes to the dataset and reclusters dataset only once. BISD algorithm provides output identical to original SNND algorithm with bounded memory overheads. We plan to generalize our BISD algorithm for other density based clustering algorithms in future.

\bibliographystyle{abbrv}
\bibliography{sample}

\end{document}